\documentclass[Physsubmission, Phys]{SciPost}
\pdfoutput=1

\hypersetup{
    colorlinks,
    linkcolor={red!50!black},
    citecolor={blue!50!black},
    urlcolor={blue!80!black}
}

\usepackage[bitstream-charter]{mathdesign}
\usepackage{slashed} %Slash Dirac notation
\usepackage{physics} %%Bra Kets
\newcommand{\ns}{{\slashed{n}}}
\newcommand{\nsb}{{\slashed{\bar{n}}}}
\newcommand{\nn}{\nonumber\\}
\newcommand{\eps}{\epsilon}
\newcommand{\gcusp}{\Gamma_{\rm cusp}}
\newcommand{\gncuspJ}{\gamma^{\rm J}}

\newcommand{\facn}{g(n)}
\newcommand{\als}{\alpha_s}
\newcommand{\taub}{\bar{\tau}}

\def\G#1{\Gamma_{#1}}
\def\g#1{\gamma_{#1}}
\def\b#1{\beta_{#1}}
\def\C#1{c_{#1}^J}
\newcommand{\km}{k_-}
\newcommand{\kp}{k_+}
\newcommand{\kT}{k_T}
\newcommand{\lm}{l_-}
\newcommand{\lp}{l_+}
\newcommand{\lT}{l_T}
\newcommand{\thl}{\theta_l}
\newcommand{\thk}{\theta_k}
\newcommand{\thkl}{\theta_{kl}}

\DeclareSymbolFont{usualmathcal}{OMS}{cmsy}{m}{n}
\DeclareSymbolFontAlphabet{\mathcal}{usualmathcal}

\makeindex

\begin{document}

\begin{center}{\Large \textbf{Automation of Beam and Jet functions at NNLO
}}\end{center}

% TODO: write the author list here. Use initials + surname format.
% Separate subsequent authors by a comma, omit comma at the end of the list.
% Mark the corresponding author with a superscript *.
\begin{center}
Guido Bell, %\textsuperscript{1},
Kevin Brune, %\textsuperscript{2}
Goutam Das\textsuperscript{$\star$}, and 
Marcel Wald%\textsuperscript{4}
\end{center}

\begin{center}
Theoretische Physik 1, 
Center for Particle Physics, 
Universit{\"a}t Siegen,\\
Walter-Flex-Strasse 3, 
57068 Siegen, Germany
\\
$\star$ \href{mailto:goutam.das@uni-siegen.de}{goutam.das@uni-siegen.de}
\\
$\dagger$~SI-HEP-2021-25,~P3H-21-069
\end{center}

\begin{center}
\today
\end{center}

% For convenience during refereeing (optional),
% you can turn on line numbers by uncommenting the next line:
%\linenumbers

\definecolor{palegray}{gray}{0.95}
\begin{center}
\colorbox{palegray}{
  \begin{tabular}{rr}
  \begin{minipage}{0.1\textwidth}
    \includegraphics[width=35mm]{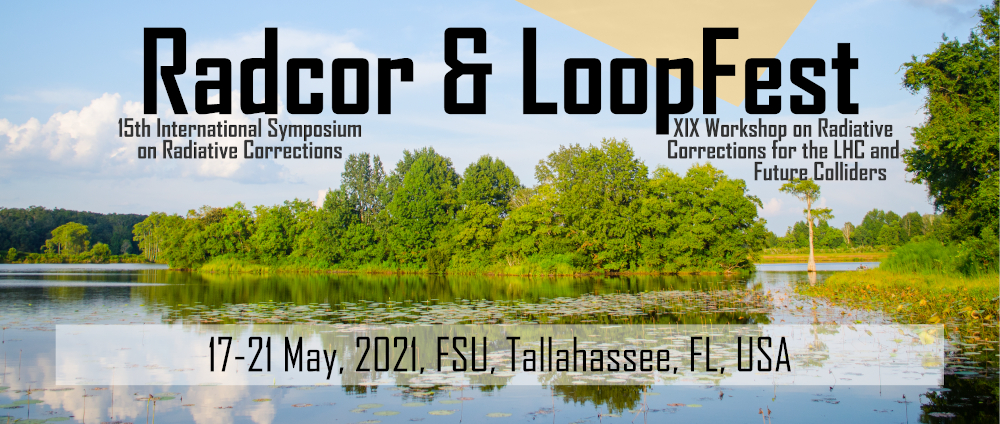}
  \end{minipage}
  &
  \begin{minipage}{0.85\textwidth}
    \begin{center}
    {\it 15th International Symposium on Radiative Corrections: \\Applications 
    of Quantum Field Theory to Phenomenology,}\\
    {\it FSU, Tallahasse, FL, USA, 17-21 May 2021} \\
%    \doi{10.21468/SciPostPhysProc.?}\\
    \end{center}
  \end{minipage}
\end{tabular}
}
\end{center}

\section*{Abstract}
{\bf
% TODO: write your abstract here.
%The abstract is in boldface, and should fit in 8 lines.
We present a novel framework to streamline the calculation of jet and beam functions 
to next-to-next-to-leading order (NNLO) in perturbation theory. By exploiting the
infra\-red behaviour of the collinear splitting functions, we factorise the singularities 
with suitable phase-space parametrisations and perform the observable-dependent 
integrations numerically. We have implemented our approach in the publicly available 
code {\tt pySecDec} and present first results for sample jet and beam functions.
}

\vspace{10pt}
\noindent\rule{\textwidth}{1pt}
\tableofcontents\thispagestyle{fancy}
\noindent\rule{\textwidth}{1pt}
\vspace{10pt}

\clearpage

%%%%%%%%%%%%%%%%%%%%%%%%%%%%%%%%%%%%%%%%%%%%%%%%%%%%%%%%%%%%%%%%%%%%%%
\section{Introduction}
%%%%%%%%%%%%%%%%%%%%%%%%%%%%%%%%%%%%%%%%%%%%%%%%%%%%%%%%%%%%%%%%%%%%%%
\label{sec:intro}

In recent years, Soft-Collinear Effective Theory (SCET) has been successful in 
describing observables at lepton as well as hadron colliders through the resummation 
of large logarithms appearing in different corners of phase space. The
resummation in the SCET framework relies on an underlying factorisation theorem
consisting of hard ($H$), beam ($B$), jet ($J$), and soft ($S$) functions,
\begin{align}
  d\sigma  = H \cdot 
            \prod_{i} B_i \otimes 
            \prod_{j} J_{j} \otimes 
            S \,.
\end{align}
All these functions can be calculated in perturbation theory order-by-order in the
strong-coupling expansion. The resummation can be performed by evolving them to a
common renormalisation scale by solving the corresponding renormalisation group
equations (RGE). The calculation of these functions often becomes challenging at 
higher orders. In particular, the jet, beam, and soft quantities depend on the
specific observable and need to be calculated on a case-by-case basis. In recent
years, there have been efforts in automating the calculation of these perturbative
ingredients. Whereas di-jet soft functions are now available to 
next-to-next-to-leading order (NNLO) for many SCET-1 and SCET-2 observables 
through the publicly available package 
{\tt SoftSERVE}~\cite{Bell:2018vaa,Bell:2018oqa,Bell:2020yzz}, an extension to  
general N-jet soft functions is currently in progress~\cite{Bell:2018mkk}. There 
have also been efforts in automating the calculation of jet and beam functions 
at NLO \cite{kbruneMS,Basdew-Sharma:2020wva}, and in this work we plan to bring
these efforts to the NNLO level, which is essential to achieve high-precision
resummations at collider processes.

%%%%%%%%%%%%%%%%%%%%%%%%%%%%%%%%%%%%%%%%%%%%%%%%%%%%%%%%%%%%%%%%%%%%%%
\section{Jet functions}
%%%%%%%%%%%%%%%%%%%%%%%%%%%%%%%%%%%%%%%%%%%%%%%%%%%%%%%%%%%%%%%%%%%%%%

The jet functions appear in SCET factorisation theorems whenever one considers
processes with coloured partons in the final state at both lepton or 
hadron colliders. In this work, we focus on quark jet functions, which are defined 
in terms of the collinear field operator 
$\chi = W^{\dagger}_{\bar n}\frac{\ns \nsb}{4} \psi$  via
\begin{align}
\label{eq:jet:definition}
  \Big{[}\frac{\ns}{2}\Big{]}_{_{\beta \alpha}} 
  J_q(\tau,\mu) = 
  \frac{1}{\pi}
  \sum_X (2\pi)^d 
  \delta\Big(Q-\sum_i k_i^-\Big)
  \delta^{(d-2)}\Big(\sum_ik_i^{\perp}\Big) 
  \bra{0} \chi_{\beta} \ket{X}
  \bra{X} \bar{\chi}_{\alpha} \ket{0}
  {\cal M}(\tau;\{k_i\}) \,, %\nn
  % &\bra{0} W_c^+(0) \frac{\ns \nsb}{4} \psi(0) 
  % \ket{X}\bra{X} \bar{\psi}(0) \frac{\nsb\ns}{4} W_c(0)  \ket{0}\,.
\end{align}
where $W_{\bar n}$ denotes a collinear Wilson line, and we introduced 
light-cone coordinates with $k_i^- = \bar n \cdot k_i$, 
$k_i^+ = n \cdot k_i$ and a transverse component $k_i^{\perp,\mu}$ satisfying  
$n \cdot k_i^{\perp}=\bar n \cdot k_i^{\perp}=0$, along with $n^2=\bar n^2=0$ and 
$n\cdot \bar n =2$. The sum over $X$ refers to the phase space of the final-state
particles and ${\cal M}(\tau;\{k_i\})$ denotes a generic measurement 
function in Laplace space, with $\tau$ being the corresponding Laplace variable 
and $\{k_i\}$ the momenta of the final-state particles.
The expansion of the bare jet functions in the strong coupling ($a_s =\als/4\pi$) 
in Laplace space can then be written as
\begin{align}
  J_q^0(\tau)
  &=
  1 +
  \sum_{k=1}^\infty\left( Z_{\alpha} a_s\right)^k
  \left(\mu^2 \taub^2\right)^{k \eps}
  J_{q}^{(k)}(\tau,\eps)\,,
  \label{eq:jet:pertexp}
\end{align} 
where $\bar\tau=\tau e^{\gamma_E}$ with $\gamma_E\simeq 0.5772$ being Euler's constant, 
and $ Z_{\alpha} = 1 - a_s \beta_0/\eps$ 
is the strong-coupling renormalisation constant in $d=4-2\eps$ dimensions 
in the $\overline{\text{MS}}$-scheme.

%%%%%%%%%%%%%%%%%%%%%%%%%%%%%%%%%%%%%%%%%%%%%%%%%%%%%%%%%%%%%%%%%%%%%%
\subsection{NLO calculation}
%%%%%%%%%%%%%%%%%%%%%%%%%%%%%%%%%%%%%%%%%%%%%%%%%%%%%%%%%%%%%%%%%%%%%%

At NLO one encounters a two-body phase space for the calculation of the jet functions 
in terms of an on-shell gluon ($k^\mu$) and a quark ($p^\mu$) momentum. However, according 
to \eqref{eq:jet:definition} the sum of their large components is constrained by the total 
jet energy $Q$ and their transverse momenta must balance each other. This leads to 
the following parametrisation (defining $\bar z=1-z$),
\begin{align}
  k_- = z Q, \qquad 
  p_- = \bar{z} Q, \qquad 
  |\vec{k}_\perp| =  |\vec{p}_\perp| = k_T, 
  \qquad \cos \theta_k = 1-2t_k \,.
\end{align} 
At NLO the collinear matrix element is proportional to the well-known 
splitting function \cite{Altarelli:1977zs},
\begin{align}
  P_{q^{*} \to gq}^{(0)}(z) 
  &=
   \frac{C_F}{z}
  \left[
    1+ \bar{z}^2 - \eps z^2
  \right] \,.
\end{align}
We then parametrise the one-emission measurement function in the form 
(see also~\cite{Bell:2018vaa,Bell:2018oqa,Bell:2020yzz}), 
\begin{align}
\label{eq:jet:measure:one-emission}
  {\cal M}_1(\tau;p,k) 
  &= 
  \exp 
  \left[ 
        -\tau k_T 
        \left( 
          \frac{k_T}{z Q}
        \right)^n
        f(z,t_k)
  \right]\,,
\end{align}
where the exponential is a result of the Laplace transformation of the momentum-space 
measurement function, and the function $f(z,t_k)$ is assumed to be 
finite and non-zero in the singular limit of the matrix element $z\to 0$. 
With this \textit{ansatz} for the generic measurement function, we arrive 
at the following master formula for the NLO quark jet functions,
\begin{align}
  J_q^{(1)} (\tau,\eps)
  =& 
  \left(\tau Q\right)^{\frac{-2n\eps}{1+n}}
  \frac{8 e^{-\gamma_E\eps}}{(1+n)\sqrt{\pi}} \,
  \frac{\Gamma\left( -\frac{2\eps}{1+n}\right)}{\Gamma\left(\frac{1}{2}-\eps\right)}
   \nn
  &\times\;\int_0^1 \!dz \;\,
  z^{-1 - \frac{2n\eps}{1+n}} \; \left[z P_{q^{*}\to gq}^{(0)}(z)\right]\;
  \int_0^1 \! d\!t_k \;\,
  (4 t_k \bar{t}_k)^{-\frac{1}{2}-\eps}\;
  f(z,t_k)^{\frac{2\eps}{1+n}}\,.
\end{align}
Notice that the $k_T$-integration is already performed, and we are thus left with two 
remaining integrations over the splitting variable $z$ and the angular variable $t_k$,
which we perform numerically. As seen above, all singularities are nicely factorised in 
this representation in terms of the gamma function and the monomial 
$z^{-1-\frac{2n\eps}{1+n}}$. The explicit dependence on the observable then enters 
through the parameter $n$ defined in~\eqref{eq:jet:measure:one-emission} and the 
function $f(z,t_k)$, which parametrises the dependence on the splitting variable 
and the azimuthal angle $\theta_k$.

%%%%%%%%%%%%%%%%%%%%%%%%%%%%%%%%%%%%%%%%%%%%%%%%%%%%%%%%%%%%%%%%%%%%%%
\subsection{NNLO calculation}
%%%%%%%%%%%%%%%%%%%%%%%%%%%%%%%%%%%%%%%%%%%%%%%%%%%%%%%%%%%%%%%%%%%%%%

The NNLO contribution to the jet functions involves two different kinds of 
contributions, $viz.$ the real-virtual (RV) and the real-real (RR) contribution. 
In the RV case, the calculation follows similarly to the NLO case since the 
phase space still consists of two particles. In particular, the measurement
function is again given by~\eqref{eq:jet:measure:one-emission} and the
matrix element is now related to the one-loop correction to the splitting 
function~\cite{Bern:1995ix,Bern:1999ry,Kosower:1999rx},
\begin{align}
  \mathbb{P}_{q^{*} \to gq}^{(1)}(z) 
  =& \frac{C_F}{z^{1-\eps}\bar{z}^{-\eps} }
  \bigg\{ ( 1+ \bar{z}^2 - \eps z^2) \bigg[ C_F+ (C_F-C_A) \,
  \Big(1-\frac{\eps^2}{1-2\eps}\Big)
  - C_A \;{}_2F_1\Big(1, -\eps; 1 - \eps; -\frac{\bar z}{z}\Big)
  \nn
  &  \hspace{7mm} +
  (C_A - 2C_F) \, {}_2F_1\Big(1, -\eps; 1 - \eps; -\frac{z}{\bar z}\Big)\bigg]
  + (C_F - C_A) \, \frac{\eps^2 \bar z(1+\bar z)}{1 - 2\eps}
  \bigg\} \,.
\end{align}
In terms of this, the master formula for the RV contribution takes the form, 
\begin{align}
J^{(2)}_{q,RV}(\tau,\eps) =& 
                \left(\tau Q\right)^{\frac{-4n\eps}{1+n}}
                \frac{ 4^{2+\eps}\pi ~e^{-2\gamma_E\eps}}{1+n}\,
                \frac{ \Gamma(-\frac{4\eps}{1+n}) \cot(\pi \eps)}
                {\eps~\Gamma(1/2 - \eps)^2}
\nn
  &\times\;
    \int_0^1 \!dz \;\,
    z^{-1-\frac{4 n \eps}{1+n}}  \; 
    \left[z ~\mathbb{P}_{q^{*}\to gq}^{(1)}(z)\right]\;
        \int_0^1 \! d\!t_k \;\,
        (4 t_k \bar{t}_k)^{-\frac{1}{2} - \eps}\;
        f(z,t_k)^{\frac{4\eps}{1+n}}\,.
\end{align}
The calculation of the RR contribution is, on the other hand, more involved
due to the complicated singularity structure of the $1\to 3$ splitting 
functions~\cite{Campbell:1997hg,Catani:1998nv,Catani:1999ss}. In particular, it turns 
out one cannot factorise all the overlapping singularities of the matrix elements with a 
single parametrisation in this case. In order to tackle the RR contribution, we then start 
from the following parametrisation of the three-body phase space,
\begin{align}\label{eq:parametrizationRR}
  a = \frac{\km \lT}{\lm \kT}, \qquad
  b = \frac{\kT}{\lT}, \qquad
  z = \frac{\km + \lm}{Q}, \qquad
  q_T = \sqrt{(\km + \lm)(\kp + \lp)}\,,
\end{align}
where $k^\mu$ and $l^\mu$ are the momenta of the emitted partons, while we again denote 
the final-state momentum of the mother quark by $p^\mu$. Here, $a$ is a measure of the 
rapidity difference of the emitted daughter particles, $b$ is the ratio of their transverse 
momenta, and $z$ and $q_T$ parametrise the dependence on their total light-cone
momenta. In addition, we also need three angles $\thk, \thl, \thkl$, which we rewrite
in terms of $t_k, t_l, t_{kl}$, similar to the NLO parametrisation discussed above. We also
find it convenient to remap the variables $a$ and $b$ to the unit hypercube, which 
automatically introduces four sectors for each contribution.  

To factorise the divergences of the RR contribution, we employ a mixed strategy 
of sector-decomposition steps, non-linear transformations 
and selector functions (details will be given in~\cite{BBDW}). We then introduce 
a similar generic measurement function for the two-emission case,
\begin{align}\label{eq:jet:measure:two-emission}
  {\cal M}_2(\tau; p,k,l) 
  &= 
  \exp 
  \left[ 
        -\tau q_T 
        \left( 
          \frac{q_T}{z Q}
        \right)^n
        {\cal F}(a,b,z,t_k, t_l, t_{kl})
  \right]\,,
\end{align}
whose exact form depends on the specific parametrisation and the divergences of the matrix 
element. While the RR contribution involves three colour structures 
($C_F^2, C_F C_A, C_F T_F n_f$), the calculation simplifies significantly for the 
$C_F T_F n_f$ piece, where the above parametrisation can be used to factorise all 
divergences. Specifically, the singularities arise in the limits $q_T \to 0$, $z \to 0 $ 
and in the overlap of $a\to1$ and $t_{kl} \to 0$ in this case. The overlapping 
divergence is a result of the configuration where the partons with momenta $k^\mu$ and 
$l^\mu$ become collinear to each other. It can be resolved with a simple substitution 
$(a, t_{kl}) \to (u,v)$ as shown in~\cite{Bell:2018vaa,Bell:2018oqa,Bell:2020yzz}.

The calculation of the $C_F^2$ and $C_F C_A$ colour structures, on the other hand, involve 
more complicated singularity patterns. Moreover, one needs to ensure that the 
function ${\cal F}$ defined in \eqref{eq:jet:measure:two-emission} stays finite and non-zero 
in the singular limits of the matrix elements, which poses additional constraints on the 
phase-space parametrisations. We have developed a strategy that satisfies these constraints
-- while properly factorising all singularities -- which requires, however, to introduce 
about a dozen different parametrisations for the $C_F^2$ and $C_F C_A$ contributions.

After factorising all divergences into monomials, one needs to perform a Laurent 
expansion to expose the divergences in the dimensional regulator $\eps$, followed 
by numerical integrations of the associated coefficients. To perform these steps, 
we use the publicly available package {\tt pySecDec} \cite{Borowka:2017idc}.
The numerical integrations are then performed with the {\tt Vegas} routine as implemented 
in the {\tt Cuba} library inside the {\tt pySecDec} framework.

%%%%%%%%%%%%%%%%%%%%%%%%%%%%%%%%%%%%%%%%%%%%%%%%%%%%%%%%%%%%%%%%%%%%%%
\subsection{Renormalisation}
%%%%%%%%%%%%%%%%%%%%%%%%%%%%%%%%%%%%%%%%%%%%%%%%%%%%%%%%%%%%%%%%%%%%%%

In Laplace space the renormalisation of the jet functions takes a multiplicative form, 
$ J_q = Z_{J_q} J_q^{0} $, and the corresponding renormalisation group equation reads,
\begin{align}
\label{eq:jet:RGE}
\frac{d}{d \ln\mu}  \; J_q(\tau,\mu)
&= \bigg[ 2 g(n) \,\gcusp(\alpha_s) \, L
+ \gncuspJ(\alpha_s) \bigg] \; J_q(\tau,\mu) \,,
\end{align}
where $g(n) = (n+1)/n$,   
$L=\ln \big( \mu\bar{\tau}/\left(Q\bar{\tau}\right)^{\frac{n}{1+n}}\big)$, and 
$\gcusp(\alpha_s)$ and $\gncuspJ(\alpha_s)$ are the cusp and non-cusp anomalous dimensions,
respectively. Expanding the anomalous dimensions in the form
$  G(\alpha_s) = \sum_{n=0}^{\infty} G_n a_s^{n+1}$, the solution of the RGE becomes 
up to two-loop order,
\begin{align}
  J_q(\tau,\mu) 
  =
  1\,+\,& a_s(\mu) \,
  \bigg\{
    \facn \, \G0 L^2
    + \g0^J L
    + \C1
  \bigg\}
  + a_s^2(\mu) \,
  \bigg\{
    \facn^2 \frac{\G0^2}{2} L^4
    + \facn \, \Big(\g0^J + \frac{2\b0}{3} \Big) \, \G0 L^3
    \nn 
    &+ \left( \facn \, (\G1 + \G0 \C1) +
        \g0^J \, \Big(\frac{\g0^J}{2} + \b0 \Big)       
      \right) L^2 
    + \bigg(\g1^J + \C1(\g0^J + 2\b0)\bigg) \,  L
    + \C2
  \bigg\}\,.
  \label{eq:jet:renormalised}
\end{align}  
The renormalisation constant $Z_{J_q}$ satisfies a similar RGE as \eqref{eq:jet:RGE} and 
the solution follows as
\begin{align}
  & Z_{J_q}(\tau,\mu) 
  =
  1\,+\, a_s(\mu) \,
  \left\{
    -\facn \frac{\G0}{2} \frac{1}{\eps^2}
    - \bigg( \facn \G0 L + \frac{\g0^J}{2}  \bigg) \, \frac{1}{\eps}
  \right\}
  + a_s^2(\mu) \,
  \bigg\{ 
    \facn^2 \frac{\G0^2}{8} \frac{1}{\eps^4}
    \nn
    & \quad + \left( 
      \facn^2 \frac{\G0^2}{2} L
      + \facn \G0 \Big( \frac{\g0^J}{4} + \frac{3\b0}{8} \Big)
      \right) \frac{1}{\eps^3} 
      + \bigg(
        \facn^2 \frac{\G0^2}{2} L^2 
        + \facn \frac{\G0}{2} \left(\g0^J + \b0 \right) L
        \nn
     &\qquad \;   
        - \frac{\facn \G1}{8} + \frac{(\g0^J)^2}{8} + \frac{\b0\g0^J}{4} 
               \bigg) \frac{1}{\eps^2} -
    \left( 
      \facn \frac{\G1}{2} L + \frac{\g1^J}{4} 
    \right) \frac{1}{\eps}
  \bigg\}\,.
  \label{eq:jet:Zfactor}
\end{align}
This form of $Z_{J_q}$ provides a strong check of our calculation for the higher poles, 
since the observable-independent anomalous dimensions $\Gamma_0$, $\Gamma_1$ and $\beta_0$ 
are all known (we use the conventions from~\cite{Bell:2018vaa,Bell:2018oqa,Bell:2020yzz}). 
On the other hand, we can use \eqref{eq:jet:renormalised} and \eqref{eq:jet:Zfactor} to extract 
the non-cusp anomalous dimensions $\gamma_0^J$ and $\gamma_1^J$ from the $1/\eps$ poles, and the
non-logarithmic coefficients $\C1$ and $\C2$ of the renormalised jet functions from the finite terms 
of the bare NNLO calculation.

%%%%%%%%%%%%%%%%%%%%%%%%%%%%%%%%%%%%%%%%%%%%%%%%%%%%%%%%%%%%%%%%%%%%%%
\subsection{Results}
%%%%%%%%%%%%%%%%%%%%%%%%%%%%%%%%%%%%%%%%%%%%%%%%%%%%%%%%%%%%%%%%%%%%%%

\begin{figure}[!t]
  \centerline{
  \includegraphics[width=0.45\textwidth,height=4.4cm]{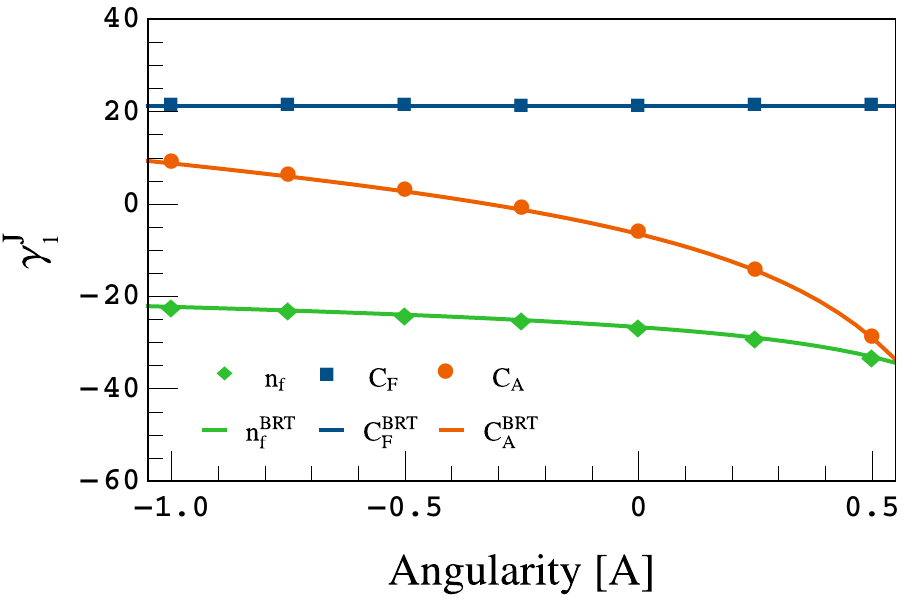}
  \hspace{5mm}
  \includegraphics[width=0.45\textwidth,height=4.4cm]{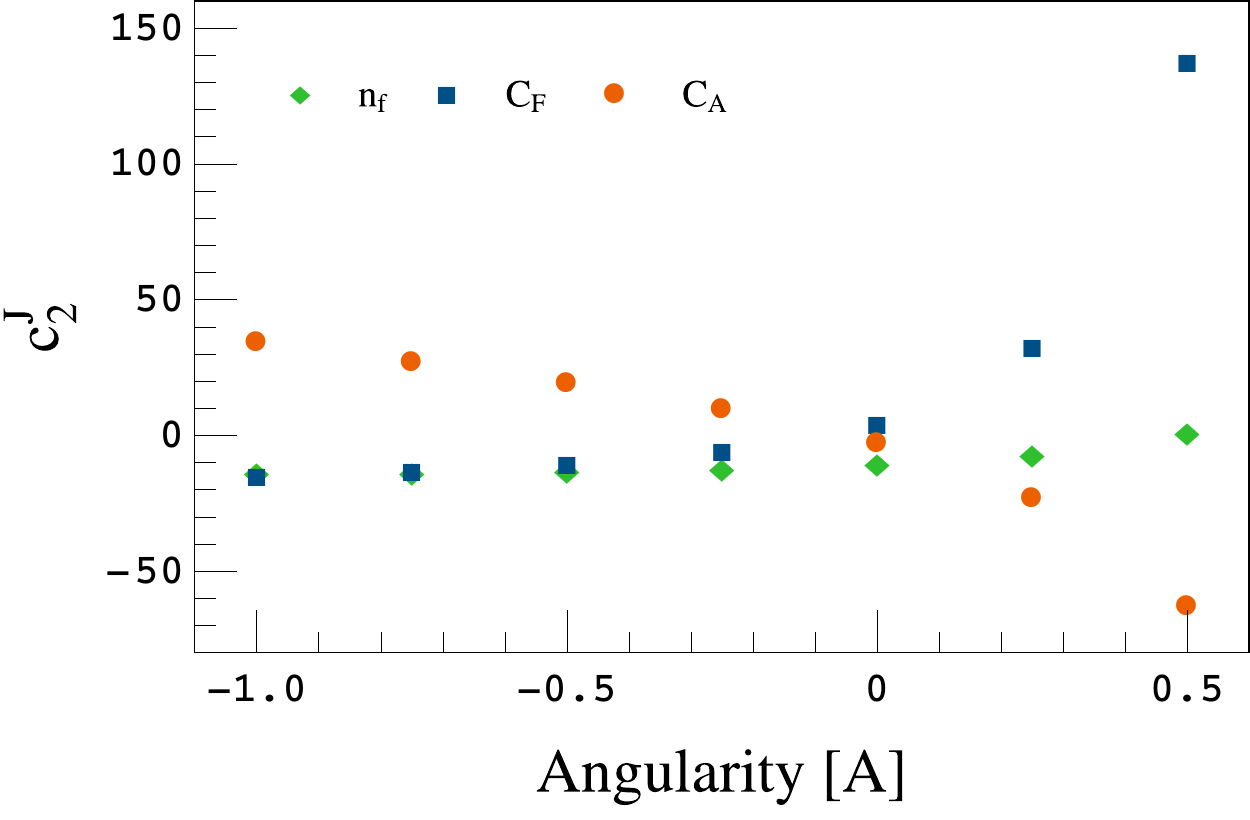}
  }
  \caption{\small The two-loop non-cusp anomalous dimensions (left) are compared against the 
  semi-analytic results (BRT) of \cite{Bell:2018vaa}. The new $\C2$ coefficients (right) are 
  obtained for different angularity values. The numerical uncertainties are too small to be 
  visible on the plot.\\[-1.5em]}
  \label{fig:angularityR}
  \end{figure}

With this setup, we computed the jet functions for the event-shape observables 
thrust and angularities. As the NLO case is trivial, we focus here on the NNLO numbers, which we 
present in the form 
$\gamma_1^J = \gamma_1^{C_F} \,C_F^2 + \gamma_1^{C_A} \,C_F C_A + \gamma_1^{n_f} \,C_F T_F n_f$,
and similarly for $\C2$. Our numbers for thrust are summarised in Table~\ref{tab:jet:thrust}, 
which show a very good agreement with the analytic results from~\cite{Becher:2006qw}.
Our uncertainty estimates, on the other hand, seem to be too conservative at present 
and require further investigations.
We also find excellent agreement for angularities for the $\gamma_1^J$ coefficients 
with the literature \cite{Bell:2018vaa}, as can be seen from the 
left panel of Figure~\ref{fig:angularityR}. On the right panel,  
we present our numbers for the $\C2$ coefficients for different angularity values, which
significantly improve the {\sc Event2} fits results from~\cite{Bell:2018gce}.

\begin{table}[!t]
  \centering
  \renewcommand\arraystretch{1.4}
\begin{tabular}{|c|c|c|}
  \hline
       &
      ~~~~~Analytic~~~~~ &
      ~~~~This work~~~~\\
  \hline
  \hline
      $\gamma_1^{n_f}$ & 
      $ -26.699 $  & 
      $ -26.699(8)$  \\
  \hline
      $\gamma_1^{C_A}$ &
      $ -6.520 $  & 
      $ -6.522(130)$  \\
  \hline
      $\gamma_1^{C_F}$ &
      $ 21.220 $  & 
      $ 21.219(119)$  \\
  \hline
  \end{tabular}
  \hspace{10mm}
    \begin{tabular}{|c|c|c|}
  \hline
       &
      ~~~~~Analytic~~~~~ &
      ~~~~This work~~~~\\
  \hline
  \hline
  $c_2^{n_f}$ &
  $-10.787$  & 
      $-10.787(15)$  \\
  \hline
    $c_2^{C_A}$ &
      $-2.165$        & 
      $-2.169(189)$    \\
  \hline
      $c_2^{C_F}$ &
      $4.655$    & 
      $4.654(146)$  \\
  \hline
  \end{tabular}
   \caption{\small{Two-loop non-cusp anomalous dimension (left) and finite term of the 
   renormalised jet function (right) for thrust. The analytic results have been extracted
   from~\cite{Becher:2006qw}.\\[-1.5em]}} 
  \label{tab:jet:thrust}
\end{table}

%%%%%%%%%%%%%%%%%%%%%%%%%%%%%%%%%%%%%%%%%%%%%%%%%%%%%%%%%%%%%%%%%%%%%%
\section{Beam functions}
%%%%%%%%%%%%%%%%%%%%%%%%%%%%%%%%%%%%%%%%%%%%%%%%%%%%%%%%%%%%%%%%%%%%%%

The beam functions are defined via proton matrix elements of collinear
field operators. In contrast to the jet functions discussed above, the beam functions 
are non-perturbative objects, which must be matched onto parton distribution 
functions (PDF) to extract the perturbative information. At the partonic level, the 
quark-to-quark beam function is defined e.g.~as
\begin{align}
\label{eq:beam:definition}
    \frac12 \, \left[\frac{\ns}{2}\right]_{\beta \alpha}
    {\cal B}_{qq}(x,\tau,\mu) = \sum_{X} 
    \delta\Big( (1-x) P_- - \sum_i k_i^- \Big)\,
    \bra{P}\bar{\chi}_{\alpha} \ket{X} 
    \bra{X}\chi_{\beta}  \ket{P} 
    {\cal M}(\tau;\{k_i\}) \,,
\end{align}
where $\ket{P}$ is now a partonic state with momentum $P^\mu$. In the general case, 
the matching onto the partonic PDF $f_{ij}$ then takes the form,
\begin{align}
    {\cal B}_{ij}(x,\tau,\mu)&=\sum_k 
    \int_x^1 \frac{dz}{z} \,
    {\cal I}_{ik}\Big(\frac{x}{z},\tau,\mu\Big) 
    ~ f_{kj}(z,\mu)\,.
\end{align}
While we use the same notation for the Laplace-space measurement function
${\cal M}(\tau;\{k_i\})$ as in the previous section, the beam functions
${\cal B}_{ij}$ and the matching kernels ${\cal I}_{ij}$ are distribution-valued
in the momentum fraction $x$. To avoid this, we perform an additional 
Mellin transformation, and we denote the corresponding quantities 
in Mellin space by $\widehat{\cal B}_{ij}$ 
and $\widehat{\cal I}_{ij}$. In Mellin space, the bare quantities have a similar 
perturbative expansion as the jet functions in \eqref{eq:jet:pertexp}.

%%%%%%%%%%%%%%%%%%%%%%%%%%%%%%%%%%%%%%%%%%%%%%%%%%%%%%%%%%%%%%%%%%%%%%
\subsection{Calculational details}
%%%%%%%%%%%%%%%%%%%%%%%%%%%%%%%%%%%%%%%%%%%%%%%%%%%%%%%%%%%%%%%%%%%%%%

The collinear matrix elements in the definition of the beam functions can be extracted 
from the well-known splitting functions using crossing symmetry. Notice that at NNLO, the 
calculation of the beam functions only involves a two-body phase space, for which we employ 
similar parametrisations as in the jet-function case discussed above. In the following,
we focus on the quark-to-quark beam function for transverse momentum resummation, which
is defined in SCET-2, and as such requires an additional rapidity regulator. 
To this end, we adopt the analytic regulator from~\cite{Becher:2011dz} 
in a symmetrised version (see also~\cite{Bell:2018vaa,Bell:2018oqa,Bell:2020yzz}),
and we follow the collinear anomaly approach~\cite{Becher:2010tm} to extract the
final matching kernels. Specifically, we write the product of soft and beam-function kernels 
in Mellin space $(N_1,N_2)$ in the form
\begin{align}
  \left[
    {\cal S}(\tau,\mu,\nu) \;
    \widehat{\cal I}_{qq}(N_1,\tau,\mu,\nu) \;
    \widehat{\cal I}_{\bar{q}\bar{q}}(N_2,\tau,\mu,\nu) 
  \right]_{q^2}
  \stackrel{\alpha=0}{\equiv}
  \left( \taub^2 q^2\right)^{-F_{q\bar{q}}(\tau,\mu)} \;
  \widehat{I}_{qq}(N_1,\tau,\mu) \;
  \widehat{I}_{\bar{q}\bar{q}}(N_2,\tau,\mu)\,,
  \label{eq:beam:refactorisation}
\end{align}
where $\alpha$ is the rapidity regulator, $\nu$ is the rapidity scale and $q^2$ refers to the 
hard scale in the problem, e.g.~the invariant mass of the Drell-Yan pair.
On the right-hand-side of \eqref{eq:beam:refactorisation}, the large rapidity logarithms are 
then resummed to 
all orders through the anomaly coefficient $F_{q\bar{q}}(\tau,\mu)$. Due to our specific
choice of the rapidity regulator, the collinear and anti-collinear matching kernels 
on the left-hand-side are furthermore symmetric under the exchange of 
$n^\mu \leftrightarrow \bar{n}^\mu$, and it is therefore sufficient to compute only one of 
them explicitly. On the other hand, we also need the soft function in the same regularisation
scheme, for which we rely on {\tt SoftSERVE}~\cite{Bell:2018vaa,Bell:2018oqa,Bell:2020yzz}.

The renormalised anomaly coefficient then satisfies the RGE
\begin{align}
  \frac{d}{d\ln \mu} \; F_{q\bar{q}}(\tau,\mu) 
  = 2\,\Gamma_{\rm cusp}(\als)\,,
\end{align} 
whose solution, up to two loops, takes the form
\begin{align}
  F_{q\bar{q}}(\tau,\mu) 
  = 
  a_s(\mu) \,\Big\{2 \G0 L + d_1 \Big\}
  + a_s^2(\mu) \,\Big\{
    2 \b0 \G0 L^2 + 2 \,(\G1  +  \b0 d_1)\, L + d_2 \,
  \Big\}
\end{align}
where $L = \ln(\mu \bar{\tau})$ and $d_i$ are the non-logarithmic terms of the renormalised anomaly coefficient. The matching kernels, on the other hand, obey the following RGE in Mellin space,
\begin{align}\label{eq:rgeI}
  \frac{d}{d\ln \mu} \;\widehat{I}_{qq}(N,\tau,\mu)
  =
  2\left[
    \Gamma_{\rm cusp}(\als) \, L +\gamma^{B}(\als)
  \right] \widehat{I}_{qq}(N,\tau,\mu)
  -2 \sum_{j} \;\widehat{I}_{qj}(N,\tau,\mu) \,\widehat{P}_{jq}(N,\mu) \,,
\end{align}
where $\gamma^B$ is the non-cusp anomalous dimension, and the sum in the last term 
of the right-hand-side runs over all partons. In this notation, $\widehat{P}_{qq}$ 
($\widehat{P}_{gq}$) is simply the Mellin-transformed splitting function  
$P_{q\to q g^*}$ ($P_{q\to g q^*}$). From the solution of \eqref{eq:rgeI}, we finally extract 
the non-logarithmic pieces $\textstyle\widehat{I}_{qq,i}(N)$ by expanding in $a_s(\mu)$
and setting $\mu=1/\bar\tau$, similar to \eqref{eq:jet:renormalised}.

%%%%%%%%%%%%%%%%%%%%%%%%%%%%%%%%%%%%%%%%%%%%%%%%%%%%%%%%%%%%%%%%%%%%%%
\subsection{Results}
%%%%%%%%%%%%%%%%%%%%%%%%%%%%%%%%%%%%%%%%%%%%%%%%%%%%%%%%%%%%%%%%%%%%%%

\begin{figure}[t]
  \centerline{
  \includegraphics[width=0.45\textwidth,height=4.4cm]{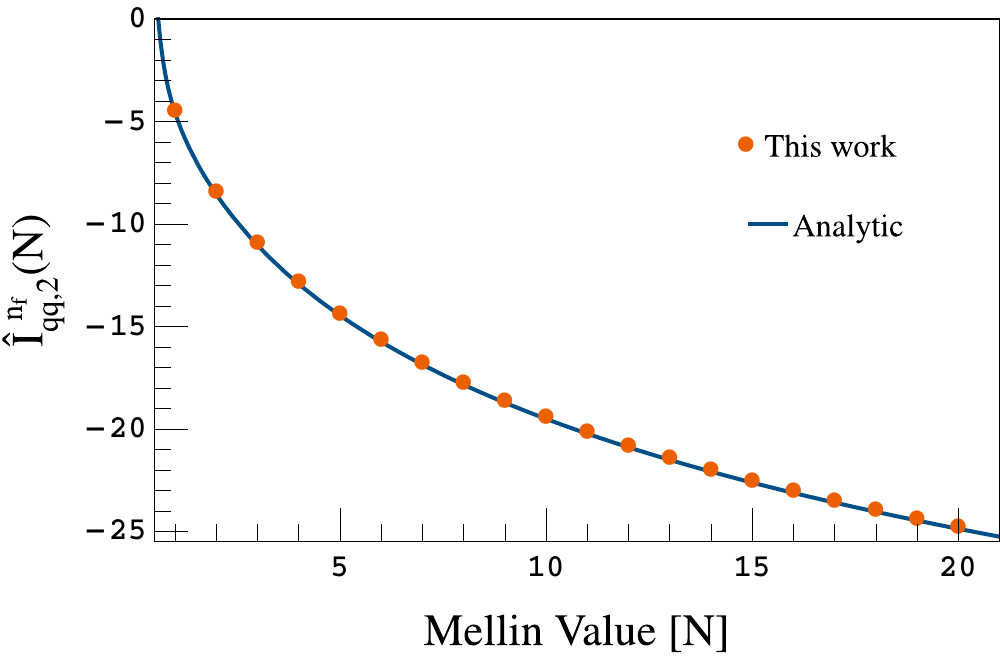}
  \includegraphics[width=0.45\textwidth,height=4.4cm]{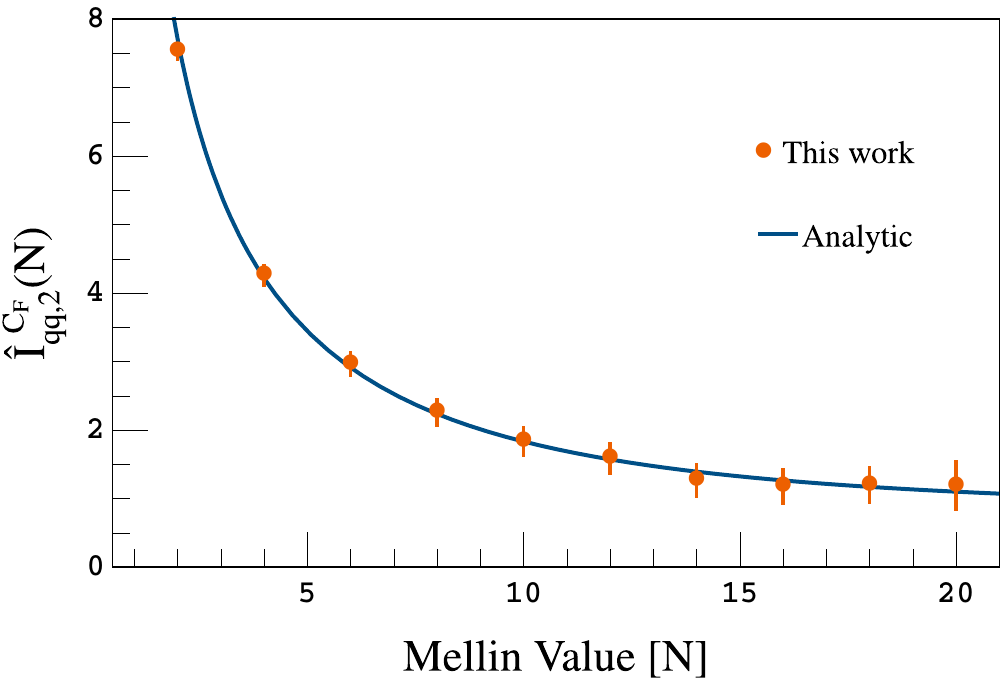}
  }
  \caption{\small Comparison of the two-loop non-logarithmic contribution to the 
  quark-to-quark matching kernel for $p_T$ resummation against the analytic 
  results of~\cite{Gehrmann:2012ze,Gehrmann:2014yya}.\\[-1.5em]  }
  \label{fig:beampT}
\end{figure}

At NNLO we have calculated all contributions except for the $C_F C_A$ colour structure 
ap\-pearing in the $P_{q \to ggq^{*}}$ splitting kernel.
\begin{table}[!t]
  \centering
  \renewcommand\arraystretch{1.4}
\begin{tabular}{|c|c|c|}
  \hline
      &
      ~~~~~Analytic~~~~~ &
      ~~~~This work~~~~\\
  \hline
  \hline
      $\gamma_1^{n_f}$ & 
      $ -11.395 $  & 
      $ -11.392(8) $  \\
  \hline
      $\gamma_1^{C_F}$ &
      $ 10.610 $  & 
      $ 10.594(40) $  \\
  \hline
  \end{tabular}
  \hspace{10mm}
    \begin{tabular}{|c|c|c|}
  \hline
      &
      ~~~~~Analytic~~~~~ &
      ~~~~This work~~~~\\
  \hline
  \hline
  $d_2^{n_f}$&
  $4.148$  & 
      $4.147(4)$  \\
  \hline
  $d_2^{C_F}$ &
      $0$        & 
      $0.024(45)$    \\
  \hline
  \end{tabular}
  \caption{\small{Two-loop non-cusp anomalous dimension (left) and anomaly coefficient (right) for $p_T$-resummation. The analytic results have been extracted from~\cite{Gehrmann:2012ze,Gehrmann:2014yya}.\\[-1.5em]}} 
  \label{tab:beam:pT}
\end{table}
As can be seen in Table~\ref{tab:beam:pT}, our results for the remaining colour structures of 
the anomaly coefficient $d_2$ as well as the non-cusp anomalous dimension $\gamma_1^B$ are 
consistent within numerical uncertainties with the analytic results from~\cite{Gehrmann:2012ze,Gehrmann:2014yya}. In Figure~\ref{fig:beampT}, we display the 
non-logarithmic contribution  to the renormalised matching kernel $\textstyle\widehat{I}_{qq,2}(N)$ 
as a function of the Mellin parameter $N$. For both colour structures computed so far,
we again find a very good agreement with the known analytic results.

%%%%%%%%%%%%%%%%%%%%%%%%%%%%%%%%%%%%%%%%%%%%%%%%%%%%%%%%%%%%%%%%%%%%%%
\section{Conclusions}
%%%%%%%%%%%%%%%%%%%%%%%%%%%%%%%%%%%%%%%%%%%%%%%%%%%%%%%%%%%%%%%%%%%%%%

We have presented a formalism to automate the calculation of two-loop jet and beam functions. 
Due to the complicated divergence structure of the underlying collinear matrix elements, 
we encountered many overlapping singularities in the RR contribution, which we disentangled
with the help of a mixed strategy based on sector decomposition, non-linear transformations 
and selector functions. We furthermore validated our setup against known results 
for the thrust jet function and the transverse-momentum-dependent beam function,
and we obtained a new prediction for the angularity jet function. While we have not yet
finished the implementation of the RR contribution in the beam-function case, we plan to look 
into more observables soon. In the longer term, we also envisage to provide a public code for the
calculation of jet and beam functions, similar to the {\tt SoftSERVE} distribution.

% \section*{Acknowledgements}
% Acknowledgements should follow immediately after the conclusion.

% TODO: include author contributions
% \paragraph{Author contributions}
% This is optional. If desired, contributions should be 
% succinctly described in a single short paragraph, using 
% author initials.

% TODO: include funding information
\paragraph{Funding information}
% Authors are required to provide funding information, 
% including relevant agencies and grant numbers with linked 
% author's initials. Correctly-provided data will be linked 
% to funders listed in the 
% \href{https://www.crossref.org/services/funder-registry/}{\sf 
% Fundref registry}.
This research was supported by the Deutsche Forschungsgemeinschaft 
(DFG, German Research Foundation) under grant 396021762 - TRR 257.

% \begin{appendix}

% \section{First appendix}
% Add material which is better left outside the main text 
% in a series of Appendices labeled by capital letters.

% \section{About references}
% Your references should start with the comma-separated 
% author list (initials + last name), the publication title 
% in italics, the journal reference with volume in bold, 
% start page number, publication year in parenthesis, 
% completed by the DOI link (linking must be implemented 
% before publication). If using BiBTeX, please use the 
% style files provided  on 
% \url{https://scipost.org/submissions/author_guidelines}. 
% If you are using our \LaTeX template, simply add
% \begin{verbatim}
% \bibliography{your_bibtex_file}
% \end{verbatim}
% at the end of your document. If you are not using our \LaTeX template, please still use our bibstyle as
% \begin{verbatim}
% \bibliographystyle{SciPost_bibstyle} % Include this style file here only if you are not using our template
\bibliography{references}

\begin{thebibliography}{10}
\providecommand{\url}[1]{\texttt{#1}}
\providecommand{\urlprefix}{URL }
\expandafter\ifx\csname urlstyle\endcsname\relax
  \providecommand{\doi}[1]{doi:\discretionary{}{}{}#1}\else
  \providecommand{\doi}{doi:\discretionary{}{}{}\begingroup
  \urlstyle{rm}\Url}\fi
\providecommand{\eprint}[2][]{\url{#2}}

\bibitem{Bell:2018vaa}
G.~Bell, R.~Rahn and J.~Talbert,
\newblock \emph{{Two-loop anomalous dimensions of generic dijet soft
  functions}},
\newblock Nucl. Phys. B \textbf{936}, 520 (2018),
\newblock \doi{10.1016/j.nuclphysb.2018.09.026},
\newblock \eprint{1805.12414}.

\bibitem{Bell:2018oqa}
G.~Bell, R.~Rahn and J.~Talbert,
\newblock \emph{{Generic dijet soft functions at two-loop order: correlated
  emissions}},
\newblock JHEP \textbf{07}, 101 (2019),
\newblock \doi{10.1007/JHEP07(2019)101},
\newblock \eprint{1812.08690}.

\bibitem{Bell:2020yzz}
G.~Bell, R.~Rahn and J.~Talbert,
\newblock \emph{{Generic dijet soft functions at two-loop order: uncorrelated
  emissions}},
\newblock JHEP \textbf{09}, 015 (2020),
\newblock \doi{10.1007/JHEP09(2020)015},
\newblock \eprint{2004.08396}.

\bibitem{Bell:2018mkk}
G.~Bell, B.~Dehnadi, T.~Mohrmann and R.~Rahn,
\newblock \emph{{Automated Calculation of N-jet Soft Functions}},
\newblock PoS \textbf{LL2018}, 044 (2018),
\newblock \doi{10.22323/1.303.0044},
\newblock \eprint{1808.07427}.

\bibitem{kbruneMS}
K.~Brune,
\newblock \emph{NLO calculation of jet and beam functions in Soft-Collinear
  Effective Theory},
\newblock Master's thesis, Universit{\"a}t Siegen (2019).

\bibitem{Basdew-Sharma:2020wva}
A.~Basdew-Sharma, F.~Herzog, S.~Schrijnder~van Velzen and W.~J. Waalewijn,
\newblock \emph{{One-loop jet functions by geometric subtraction}},
\newblock JHEP \textbf{10}, 118 (2020),
\newblock \doi{10.1007/JHEP10(2020)118},
\newblock \eprint{2006.14627}.

\bibitem{Altarelli:1977zs}
G.~Altarelli and G.~Parisi,
\newblock \emph{{Asymptotic Freedom in Parton Language}},
\newblock Nucl. Phys. B \textbf{126}, 298 (1977),
\newblock \doi{10.1016/0550-3213(77)90384-4}.

\bibitem{Bern:1995ix}
Z.~Bern and G.~Chalmers,
\newblock \emph{{Factorization in one loop gauge theory}},
\newblock Nucl. Phys. B \textbf{447}, 465 (1995),
\newblock \doi{10.1016/0550-3213(95)00226-I},
\newblock \eprint{hep-ph/9503236}.

\bibitem{Bern:1999ry}
Z.~Bern, V.~Del~Duca, W.~B. Kilgore and C.~R. Schmidt,
\newblock \emph{{The infrared behavior of one loop QCD amplitudes at
  next-to-next-to leading order}},
\newblock Phys. Rev. D \textbf{60}, 116001 (1999),
\newblock \doi{10.1103/PhysRevD.60.116001},
\newblock \eprint{hep-ph/9903516}.

\bibitem{Kosower:1999rx}
D.~A. Kosower and P.~Uwer,
\newblock \emph{{One loop splitting amplitudes in gauge theory}},
\newblock Nucl. Phys. B \textbf{563}, 477 (1999),
\newblock \doi{10.1016/S0550-3213(99)00583-0},
\newblock \eprint{hep-ph/9903515}.

\bibitem{Campbell:1997hg}
J.~M. Campbell and E.~W.~N. Glover,
\newblock \emph{{Double unresolved approximations to multiparton scattering
  amplitudes}},
\newblock Nucl. Phys. B \textbf{527}, 264 (1998),
\newblock \doi{10.1016/S0550-3213(98)00295-8},
\newblock \eprint{hep-ph/9710255}.

\bibitem{Catani:1998nv}
S.~Catani and M.~Grazzini,
\newblock \emph{{Collinear factorization and splitting functions for
  next-to-next-to-leading order QCD calculations}},
\newblock Phys. Lett. B \textbf{446}, 143 (1999),
\newblock \doi{10.1016/S0370-2693(98)01513-5},
\newblock \eprint{hep-ph/9810389}.

\bibitem{Catani:1999ss}
S.~Catani and M.~Grazzini,
\newblock \emph{{Infrared factorization of tree level QCD amplitudes at the
  next-to-next-to-leading order and beyond}},
\newblock Nucl. Phys. B \textbf{570}, 287 (2000),
\newblock \doi{10.1016/S0550-3213(99)00778-6},
\newblock \eprint{hep-ph/9908523}.

\bibitem{BBDW}
G.~Bell, K.~Brune, G.~Das and M.~Wald,
\newblock to appear .

\bibitem{Borowka:2017idc}
S.~Borowka, G.~Heinrich, S.~Jahn, S.~P. Jones, M.~Kerner, J.~Schlenk and
  T.~Zirke,
\newblock \emph{{pySecDec: a toolbox for the numerical evaluation of
  multi-scale integrals}},
\newblock Comput. Phys. Commun. \textbf{222}, 313 (2018),
\newblock \doi{10.1016/j.cpc.2017.09.015},
\newblock \eprint{1703.09692}.

\bibitem{Becher:2006qw}
T.~Becher and M.~Neubert,
\newblock \emph{{Toward a NNLO calculation of the $\bar B \to X_s \gamma$ decay
  rate with a cut on photon energy. II. Two-loop result for the jet function}},
\newblock Phys. Lett. B \textbf{637}, 251 (2006),
\newblock \doi{10.1016/j.physletb.2006.04.046},
\newblock \eprint{hep-ph/0603140}.

\bibitem{Bell:2018gce}
G.~Bell, A.~Hornig, C.~Lee and J.~Talbert,
\newblock \emph{{$e^+ e^-$ angularity distributions at NNLL$^\prime$
  accuracy}},
\newblock JHEP \textbf{01}, 147 (2019),
\newblock \doi{10.1007/JHEP01(2019)147},
\newblock \eprint{1808.07867}.

\bibitem{Becher:2011dz}
T.~Becher and G.~Bell,
\newblock \emph{{Analytic Regularization in Soft-Collinear Effective Theory}},
\newblock Phys. Lett. B \textbf{713}, 41 (2012),
\newblock \doi{10.1016/j.physletb.2012.05.016},
\newblock \eprint{1112.3907}.

\bibitem{Becher:2010tm}
T.~Becher and M.~Neubert,
\newblock \emph{{Drell-Yan Production at Small $q_T$, Transverse Parton
  Distributions and the Collinear Anomaly}},
\newblock Eur. Phys. J. C \textbf{71}, 1665 (2011),
\newblock \doi{10.1140/epjc/s10052-011-1665-7},
\newblock \eprint{1007.4005}.

\bibitem{Gehrmann:2012ze}
T.~Gehrmann, T.~Luebbert and L.~L. Yang,
\newblock \emph{{Transverse parton distribution functions at
  next-to-next-to-leading order: the quark-to-quark case}},
\newblock Phys. Rev. Lett. \textbf{109}, 242003 (2012),
\newblock \doi{10.1103/PhysRevLett.109.242003},
\newblock \eprint{1209.0682}.

\bibitem{Gehrmann:2014yya}
T.~Gehrmann, T.~Luebbert and L.~L. Yang,
\newblock \emph{{Calculation of the transverse parton distribution functions at
  next-to-next-to-leading order}},
\newblock JHEP \textbf{06}, 155 (2014),
\newblock \doi{10.1007/JHEP06(2014)155},
\newblock \eprint{1403.6451}.

\end{thebibliography}
% \bibstyle{SciPost_bibstyle}
% \end{verbatim}
% in order to simplify the production of your paper.
% \end{appendix}

% TODO:
% Provide your bibliography here. You have two options:

% FIRST OPTION - write your entries here directly, following the example below, including Author(s), Title, Journal Ref. with year in parentheses at the end, followed by the DOI number.
%\begin{thebibliography}{99}
%\bibitem{1931_Bethe_ZP_71} H. A. Bethe, {\it Zur Theorie der Metalle. i. Eigenwerte und Eigenfunktionen der linearen Atomkette}, Zeit. f{\"u}r Phys. {\bf 71}, 205 (1931), \doi{10.1007\%2FBF01341708}.
%\bibitem{arXiv:1108.2700} P. Ginsparg, {\it It was twenty years ago today... }, \url{http://arxiv.org/abs/1108.2700}.
%\end{thebibliography}

% \bibliographystyle{SciPost_bibstyle} % Include this style file here only if you are not using our template
% \bibliography{references}

\nolinenumbers

\end{document}